# Interfacial Stress Transfer in a Graphene Monolayer Nanocomposite

L. Gong[1], I. A. Kinloch[1], R. J. Young[1],*, I. Riaz[2], R. Jalil[2] & K. S. Novoselov[2],†

[1] School of Materials, University of Manchester, Grosvenor Street, Manchester, M1 7HS, UK
[2] School of Physics & Astronomy, University of Manchester, Oxford Road, Manchester, M13 9PL, UK

*E-mail: robert.young@manchester.ac.uk
[†] E-mail: kostya@manchester.ac.uk

Graphene is one of the stiffest known materials, with a Young's modulus of 1 TPa, making it an ideal candidate for use as a reinforcement in high-performance composites. However, being a one-atom thick crystalline material, graphene poses several fundamental questions: (1) can decades of research on carbon-based composites be applied to such an ultimately-thin crystalline material? (2) is continuum mechanics used traditionally with composites still valid at the atomic level? (3) how does the matrix interact with the graphene crystals and what kind of theoretical description is appropriate? We have demonstrated unambiguously that stress transfer takes place from the polymer matrix to monolayer graphene, showing that the graphene acts as a reinforcing phase. We have also modeled the behavior using shear-lag theory, showing that graphene monolayer nanocomposites can be analyzed using continuum mechanics. Additionally, we have been able to monitor stress transfer efficiency and breakdown of the graphene/polymer interface.

Since graphene was first isolated in 2004 [1,2] the majority of the research effort has concentrated upon its electronic properties aimed at applications such as in electronic devices. [3,4] A recent study has investigated the elastic mechanical properties of monolayers of graphene using nanoindentation by atomic force microscopy. [5] It was shown that the material has a Young's modulus of the order of 1 TPa and an intrinsic strength of around 130 GPa, making it the strongest material ever measured. Such impressive mechanical properties have made graphene an obvious candidate for use in high-performance polymer-based composites. Carbon nanotubes are under active investigation as reinforcements in nanocomposites [6,7] and it is well established that platelet reinforcements such as exfoliated nanoclays [8,9] can be employed as additives to enhance the mechanical and other properties of polymers. Recently it has been demonstrated that polymer-based nanocomposites with chemically-treated graphene oxide as a reinforcement show dramatic improvements in both electronic [10] and mechanical [11] properties (thus a 30 K increase in the glass transition temperature is achieved for only a 1% loading by weight of the chemically-treated graphene oxide in a poly(methyl methacrylate) matrix). Issues that arise in these nanocomposites systems include the difficulty of dispersion of the reinforcing phases and stress transfer at the interface between the dispersed phase and the polymer matrix.

It is now well established that Raman spectroscopy can be used to follow stress transfer in a variety of composites reinforced with carbon–based materials such as carbon fibres [12,13] and single- and double-walled carbon nanotubes. [14-16] Such reinforcements have well-defined Raman spectra and their Raman bands are found to shift with stress which enables stress-transfer to be monitored between the matrix and reinforcing phase. Moreover, a universal calibration has been established between the rate of shift of the G' carbon Raman bands with strain [14] that allows the effective Young's modulus of the carbon reinforcement to be estimated. Recent studies have shown that since the Raman scattering from these carbon-based materials is resonantly enhanced then strong well-defined spectra can be obtained from very small amounts of the carbon materials, for example individual carbon nanotubes either isolated on a substrate [17] or debundled and isolated within polymer nanofibers. [18,19]

Raman spectroscopy has also been employed to characterise the structure and deformation of graphene. It has been demonstrated that the technique can be used to determine the number of layers in graphene films [20] (see also Supporting Information [21]). Graphene monolayers have characteristic spectra in which the G' band (also termed the 2D band) can be fitted with a single peak, whereas the G' band in bilayers is made up of 4 peaks [20], which is a consequence of the difference between the electronic structure of the two type of samples. Several recent papers have established that the Raman bands of monolayer graphene shift during deformation. [22-25] In these reports the graphene has been deformed in tension by either stretching [22,23] or compressing [24] it on a PDMS substrate [22] or a PMMA beam. [23,24] It is also found that the G band both shifts to lower wavenumber in tension and undergoes splitting. The G' band undergoes a shift in excess of -50 $cm^{-1}$/% strain which is consistent with it having a Young's modulus of over 1 TPa [14]. A recent study [25] of graphene subjected to hydrostatic pressure has shown that the Raman bands shift to higher wavenumber for this mode of deformation and that the behavior can be predicted from knowledge of the band shifts in uniaxial tension.

In this present study we have used Raman spectroscopy to monitor stress transfer in a model composite consisting of a thin polymer matrix layer and a mechanically–cleaved single graphene monolayer using the stress-sensitivity of the graphene G' band. An optical micrograph of the specimen is shown in Figure 1(a) where the approximately diamond-shaped 12 μm × 30 μm graphene monolayer is indicated and Figure 1(b) shows a schematic diagram of the specimen.

Raman spectra were obtained initially from the middle of the monolayer and Figure 2(a) shows the position of the G' band before deformation, at 0.7 % strain and then unloaded. It can be seen from Figure 2(b) that there is a large stress-induced shift of the G' band. There was a linear shift of the band up to 0.4 % strain when the stepwise deformation was halted to map the strain across the monolayer. It was then loaded up to 0.5 % and 0.6 % strain when further mapping was undertaken and finally the specimen was unloaded from 0.7 % strain. It can be seen that there was some relaxation in the specimen following each of the mapping stages so that the band shifts became irregular. In addition, the slope of the unloading line from the highest strain level is significantly higher than that of the loading line. The slope of the unloading line is ~ -60 cm$^{-1}$/% strain, similar to the behavior found for the deformation of a free-standing monolayer on a substrate. [22,23] Moreover, the G' band position after unloading is at a higher wavenumber than before loading. This behavior is consistent with the graphene undergoing slippage in the composite during the initial tensile deformation and then becoming subjected to in-plane compression on unloading.

Mapping the local strain in along a carbon fiber in a polymer matrix allows the level of adhesion between the fiber and matrix to be evaluated. [12,13] In a similar way mapping the strain across the graphene monolayer enables stress transfer from the polymer to the graphene to be followed. Figure 3 shows the local strain in the graphene monolayer determined from the stress-induced Raman band shifts at 0.4 % matrix strain. The laser beam in the spectrometer was focused to a spot around 2 μm which allows a spatial resolution of the order of 1 μm on the monolayer by taking overlapping measurements. Figure 3(a) shows the variation of axial strain across the monolayer in the direction parallel to the strain axis. It can be seen that the strain builds up from the edges and is constant across the middle of the monolayer where the strain in the monolayer equals the applied matrix strain (0.4 %). This is completely analogous to the situation of a single discontinuous fiber in a model composite when there is good bonding between the fiber and matrix. [12,13] This behavior has been analyzed using the well-established shear-lag theory [27-29] where it is assumed that there is elastic stress transfer from the matrix to the fiber through a shear stress at the fiber-matrix interface. It is relatively easy to modify the analysis for platelet rather than fiber reinforcement (see Supporting Information [21]). It is predicted from shear-lag analysis for the platelet that for a given level of matrix strain, $e_m$, the variation of strain in the graphene flake, $e_f$, with position, $x$, across the monolayer will be of the form

$$e_f = e_m \left[ 1 - \frac{\cosh\left(ns\frac{x}{l}\right)}{\cosh(ns/2)} \right] \quad (1)$$

where

$$n = \sqrt{\frac{2G_m}{E_f}\left(\frac{t}{T}\right)} \qquad (2)$$

and $G_m$ is the matrix shear modulus, $E_f$ is the Young's modulus of the graphene flake, $l$ is the length of the graphene flake in the $x$ direction, $t$ is the thickness of the graphene, $T$ is the total resin thickness and $s$ is the aspect ratio of the graphene ($l/t$) in the $x$ direction. The parameter $n$ is accepted widely as an effective measure of the interfacial stress transfer efficiency, so $ns$ depends on both the morphology of the graphene flake and the degree of interaction it has with the matrix. The curve in Figure 3(a) is a fit of Equation (1) to the experimental data using the parameter $ns$ as the fitting variable. A reasonable fit was found for $ns \sim 20$ at $e_m = 0.4$ % (see Supporting Information [21]) showing that the interface between the polymer and graphene remained intact at this level of strain and that the behavior could be modeled using the shear-lag approach.

The variation of shear stress, $\tau_i$, at the polymer-graphene interface is given by (see Supporting Information [21])

$$\tau_i = nE_f e_m \frac{\sinh\left(ns\frac{x}{l}\right)}{\cosh(ns/2)} \qquad (3)$$

and the maximum value of $\tau_i$ at the edges of the sheet for $ns = 20$ is found to be ~2.3 MPa (see Supporting Information [21]).

Equation (1) shows that the distribution of strain in the graphene monolayer in the $x$ direction in the elastic case depends upon length of the monolayer, $l$. It can be seen from Figure 1 that the flake tapers to a point in the $y$ direction and so the axial strain in the middle of the monolayer was also mapped along the $y$ direction (see Supporting Information [21]). The strain is fairly constant along most of the monolayer but falls to zero at the tip of the flake, $y = 0$. The distribution of axial graphene strain in the middle of the monolayer at $e_m = 0.4$ % was determined using Equation (1), taking into account the changing width by varying $l$ (and hence $s$). It was found that there is also excellent agreement between the measured and predicted variation of fiber strain with position on the monolayer using $ns = 20$, validating the use of the shear lag analysis (see Supporting Information [21]).

When the matrix strain was increased to $e_m = 0.6$ % a different distribution of axial strain in the graphene monolayer was obtained as shown in Figure 3(b). In this case there appears to be an approximately linear variation of the graphene strain from the edges to the centre of the monolayer up to 0.6 % strain ( $= e_m$) and a dip in the middle down to around 0.4 % strain. In this case it appears that the interface between the graphene and polymer has failed and stress transfer is taking place through interfacial friction. [29] The strain in the graphene does not fall to zero in the middle of the flake, however, showing that the flake remains intact unlike the behavior of carbon fibers undergoing fracture in the fragmentation test. [12,13] The interfacial shear stress, $\tau_i$, in this case can be determined from the slope of the lines in Figure 3(b) using the force balance equation (see Supporting Information [21])

$$\frac{de_f}{dx} = -\frac{\tau_i}{E_f t} \quad (4)$$

which gives an interfacial shear stress in the range 0.3-0.8 MPa for the lines of different slope.

There are important implications from this study for the use of graphene as a reinforcement in nanocomposites. The quality of fiber reinforcement is often described in terms of the 'critical length', $l_c$ – the parameter is small for strong interfaces and is defined as 2× the distance over which the strain rises from the fiber ends to the plateau level. [29] It can be seen from Figure 3(a) that the strain rises to about 90 % of the plateau value over about 1.5 μm from the edge of the flake making the critical length of the graphene reinforcement of the order of 3 μm. It is generally thought that in order to obtain good fiber reinforcement the fiber length should be ~$10l_c$. Hence, relatively large graphene flakes (>30 μm) will be needed before efficient reinforcement can take place. One process for efficiently exfoliating graphene to single layers reported recently produced monolayers of no larger than a few microns across. [30,31] The relatively poor level of adhesion between the graphene and polymer matrix is also reflected in the low level of interfacial shear stress, $\tau_i$, determined - carbon fibers composites have values of $\tau_i$ an order of magnitude higher (~ 20-40 MPa). [12,13] However, in the graphene composite, interfacial stress transfer will only be taking place though van der Waals bonding across an atomically smooth surface. The efficiency of reinforcement is also reflected in the value of the parameter $ns$ (= 20) in the shear lag analysis used to fit the experimental data. Since the graphene is so thin, the aspect ratio $s$ will be large (12 µm/0.35 nm = $3.5 \times 10^4$) making $n$ small ($6 \times 10^{-4}$). This value of $n$ is a factor of 4 smaller than that determined by putting the values of $G_m$ ~ 1 GPa, $E_f$ ~ 1 TPa and $t/T$ (~ 0.35 nm/100 nm) into Equation (2) ($n$ ~ $2.6 \times 10^{-3}$), showing a possible limitation of the shear-lag analysis. [28] Nevertheless, the parameter $n$ determined experimentally can be employed to monitor the efficiency of stress transfer across the graphene-polymer interface, which in this case appears to be less than ideal.

This present study has important implications for the use of graphene as a reinforcement in composites. As well as demonstrating for the first time that it is possible to map the deformation of graphene monolayer in a polymer composite using Raman spectroscopy, a number of other issues also arise. Firstly, it is quite remarkable that a spectrum can be obtained from a reinforcement only one atom thick, allowing the mechanics of nano-reinforcement to be probed directly. Secondly, it appears that the continuum mechanics approach is also valid at the atomic level - a question widely asked in the field of nanocomposites - and that the composite micromechanics developed for the case of fiber reinforcement is also valid at the atomic level for graphene monolayers. We expect that our technique will be used widely in the evaluation of graphene composites. This present study has concentrated upon pristine, untreated graphene. Chemical modification [10] of the surface or edges may significantly strengthen the interface between the graphene and a polymer, reducing the critical length and increasing $n$. Our technique should allow the effect of chemical modification to be evaluated. Moreover, if graphene is to be used in devices in electronic circuits, it will have to be encapsulated within a polymer. The technique will also allow the effect of encapsulation upon residual stresses in the material to be probed.

*Experimental*

The specimen was prepared using a 5 mm thick poly(methyl methacrylate) beam spin-coated with 300 nm of SU-8 epoxy resin. The graphene, produced by the mechanical cleaving of graphite, was deposited on the surface of the SU-8. This method produced graphene with a range of different numbers of layers and the monolayers were identified both optically [26] and using Raman spectroscopy (see Supporting Information [21]). A thin 50 nm layer of PMMA was then spin-coated on top of the beam so that the graphene remained visible when sandwiched between the two coated polymer layers as shown in Figure 1(a).

The PMMA beam was deformed in 4-point bending and the strain monitored using a strain gauge attached to the beam surface. A well-defined Raman spectrum could be obtained through the PMMA coating using a low-power HeNe laser (1.96 eV and < 1 mW at the sample in a Renishaw 2000 spectrometer) and the deformation of the graphene in the composite was followed from the shift of the G' band [22-25]. The laser beam polarization was always parallel to the tensile axis.

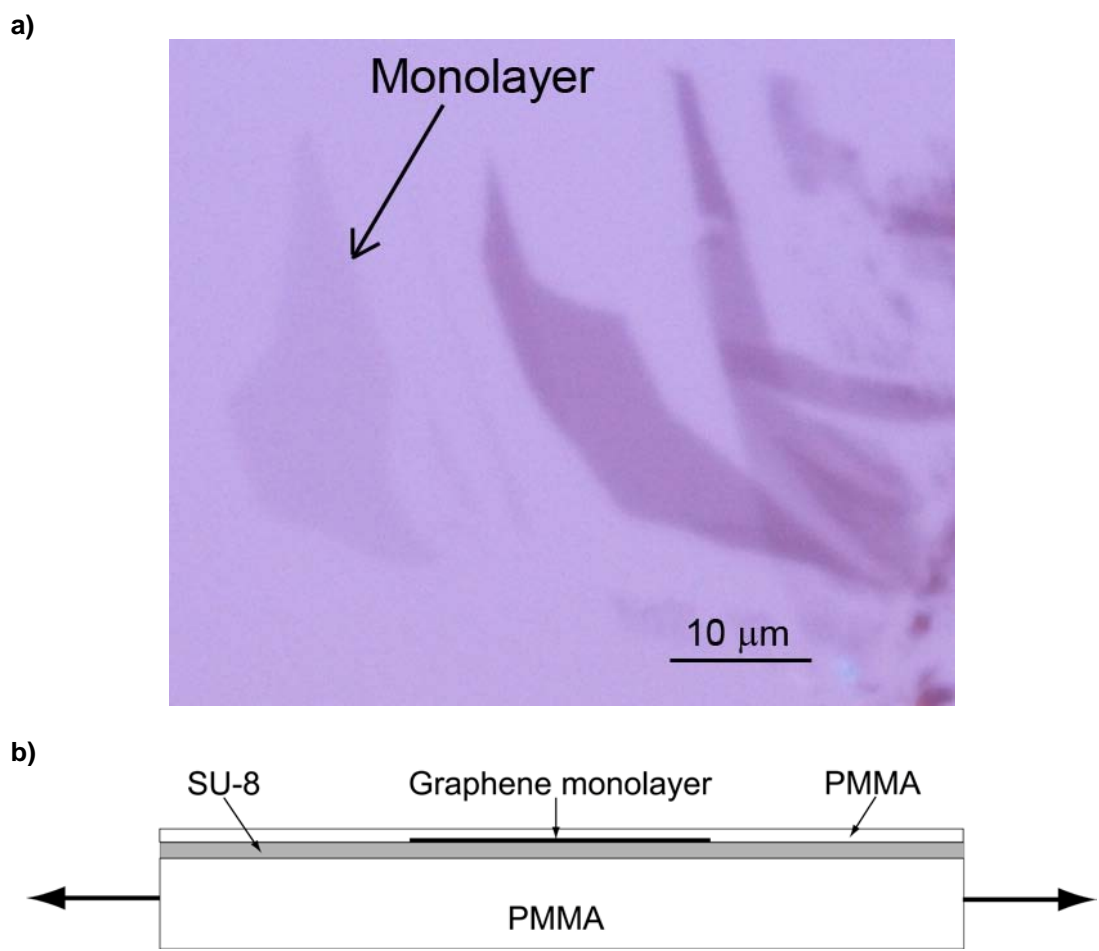

**Figure 1.** Single monolayer graphene composite. a) Optical micrograph showing the monolayer graphene flake investigated. b) Schematic diagram (not to scale) of a section through the composite

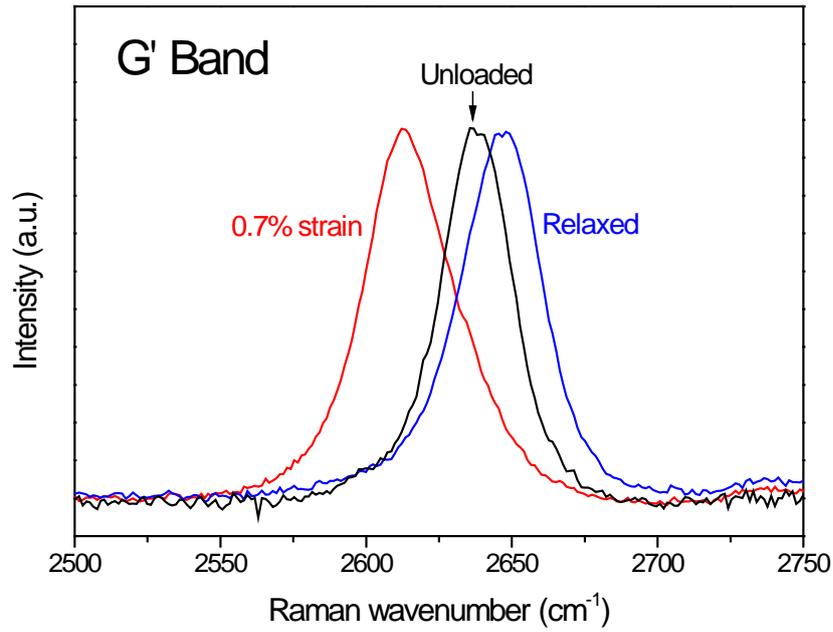

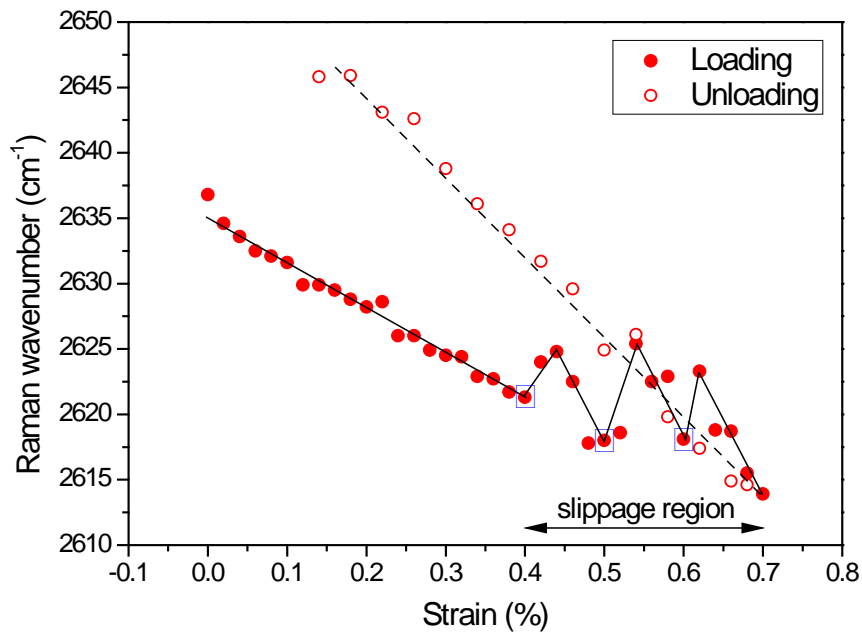

**Figure 2.** Shifts of the Raman G' band during loading and unloading of the monolayer graphene composite. a) Change in the position of the G' band with deformation. b) Shift of the G' band peak position as a function of strain. (The blue circles indicate where the loading was halted to map the strain across the flake).

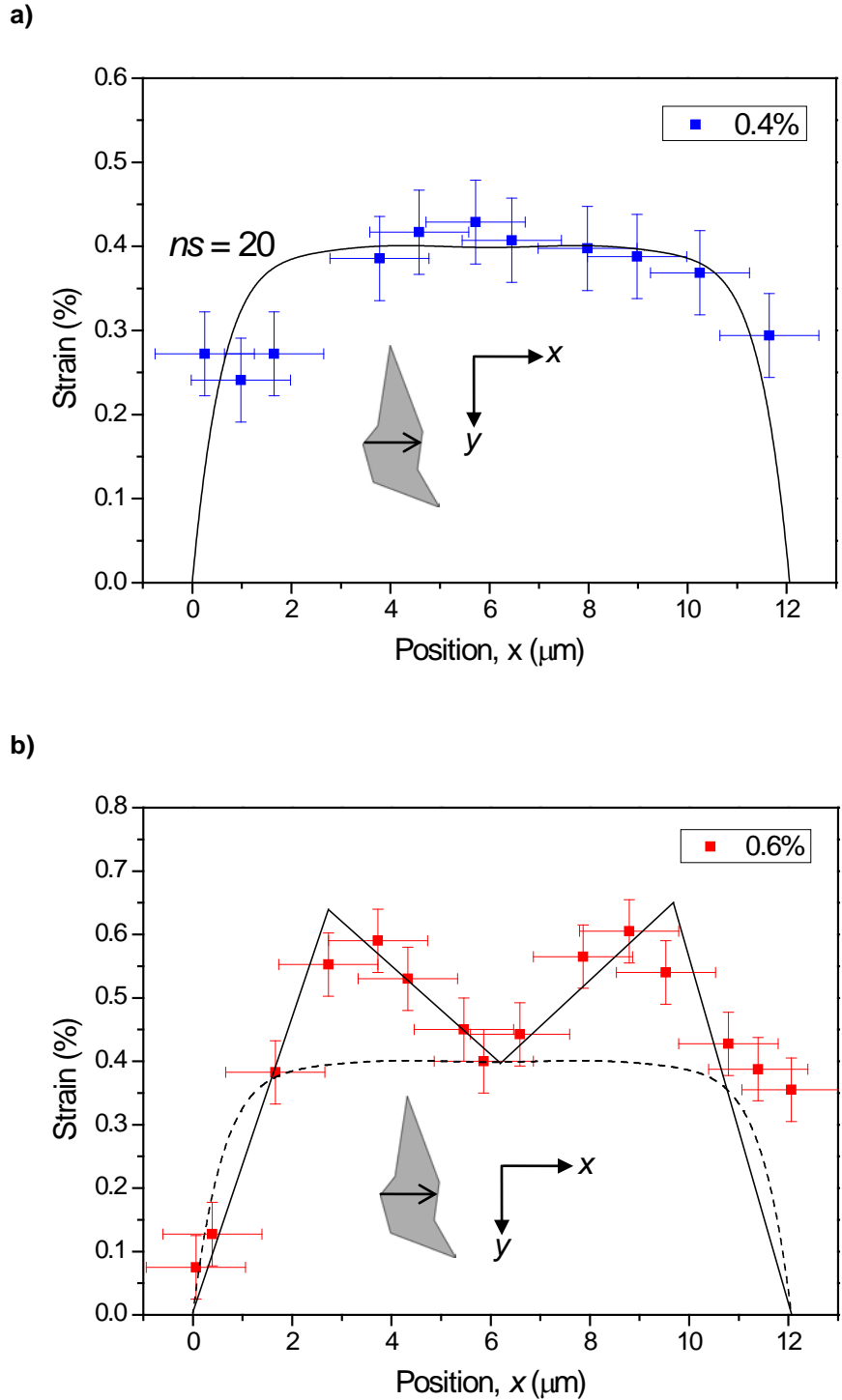

**Figure 3.** Distribution of strain in the graphene in the direction of the tensile axis (*x*) across a single monolayer. a) Variation of axial strain with position across the monolayer in the *x*-direction at 0.4% matrix strain (The curve fitted to the data is Equation (1)). b) Variation of graphene strain in the direction of the tensile axis (*x*) across a single monolayer at 0.6% matrix strain. (The solid lines are fitted to the data to guide the eye. The dashed curve is the shear-lag fit to the data in Figure 3(a) at 0.4% strain.)

# Interfacial stress transfer in a graphene monolayer nanocomposite

L. Gong[1], I. A. Kinloch[1], R. J. Young*[1], I. Riaz[2], R. Jalil[2], and K. S. Novoselov[†2]

**SI.1 Characterisation of the Graphene using Raman Spectroscopy**[S1]

Raman spectroscopy has been employed to follow the deformation of the graphene in the polymer composite. Fig. SI.1 shows that the technique can also be used to differentiate between flakes of graphene with different numbers of layers.

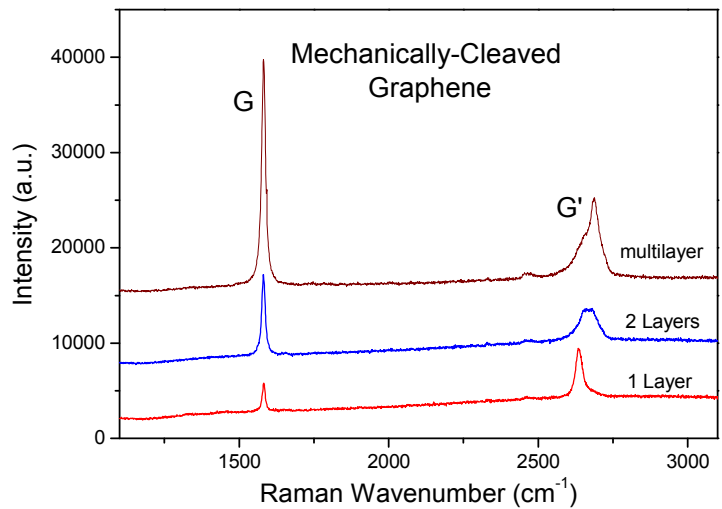

Figure SI.1 Raman spectra for different layer flakes of graphene

**SI.2 Shear Lag Analysis for a Graphene Single Monolayer**[S2, S3]

In the case of discontinuous graphene flakes reinforcing a composite matrix, stress transfer from the matrix to the flake is assumed to take place through a shear stress at the flake/matrix interface as shown in Fig. SI.2. Before deformation parallel lines perpendicular to the flake can be drawn before deformation from the matrix through the flake. When the system is subjected to axial stress, $\sigma_1$, parallel to the flake axis, the lines become distorted since the Young's modulus of the matrix is much less than that of the flake. This induces a shear stress at the flake/matrix interface. The axial stress in the flake will build up from zero at the flake ends to a maximum value in the middle of the flake. The uniform strain assumption means that, if the flake is long enough, in the middle of the flake the strain in the flake equals that in the matrix. Since the flakes have a much higher Young's modulus it means that the flakes carry most of the stress in the composite.

[1] School of Materials, University of Manchester, Grosvenor Street, Manchester, M1 7HS, UK
[2] School of Physics and Astronomy, University of Manchester, Oxford Road, Manchester, M13 9PL, UK
*E-mail: robert.young@manchester.ac.uk
†E-mail: kostya@manchester.ac.uk

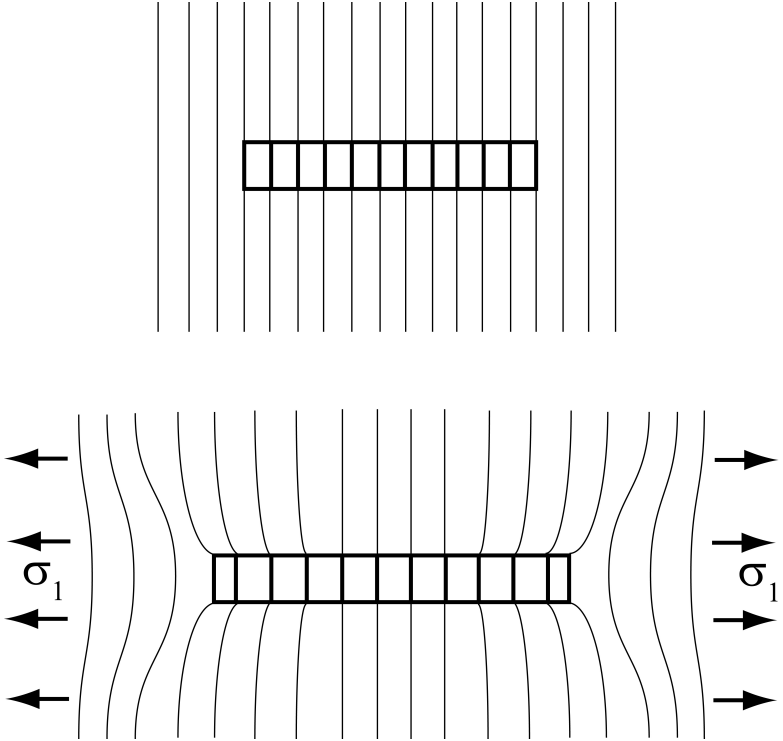

Figure SI.2 Deformation patterns for a discontinuous flake in a polymer matrix

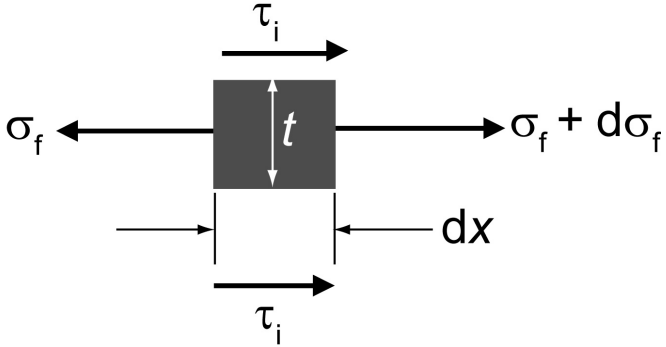

Figure SI.3 Balance of stresses acting on an element of length, d$x$, of the flake of thickness, $t$, in the composite

The relationship between the interfacial shear stress, $\tau_i$, near the flake ends and the flake stress, $\sigma_f$, can be determined by using a force balance of the shear forces at the interface and the tensile forces in a flake element as shown in Fig. SI3.

The main assumption is that the forces due to the shear stress at the interface, $\tau_i$, is balanced by the force due to the variation of axial stress in the flake, $d\sigma_f$, such that if the element shown in Fig. SI.3 is of unit width

$$\tau_i dx = -t d\sigma_f \qquad (SI.1)$$

and so

$$\frac{d\sigma_f}{dx} = -\frac{\tau_i}{t} \qquad (SI.2)$$

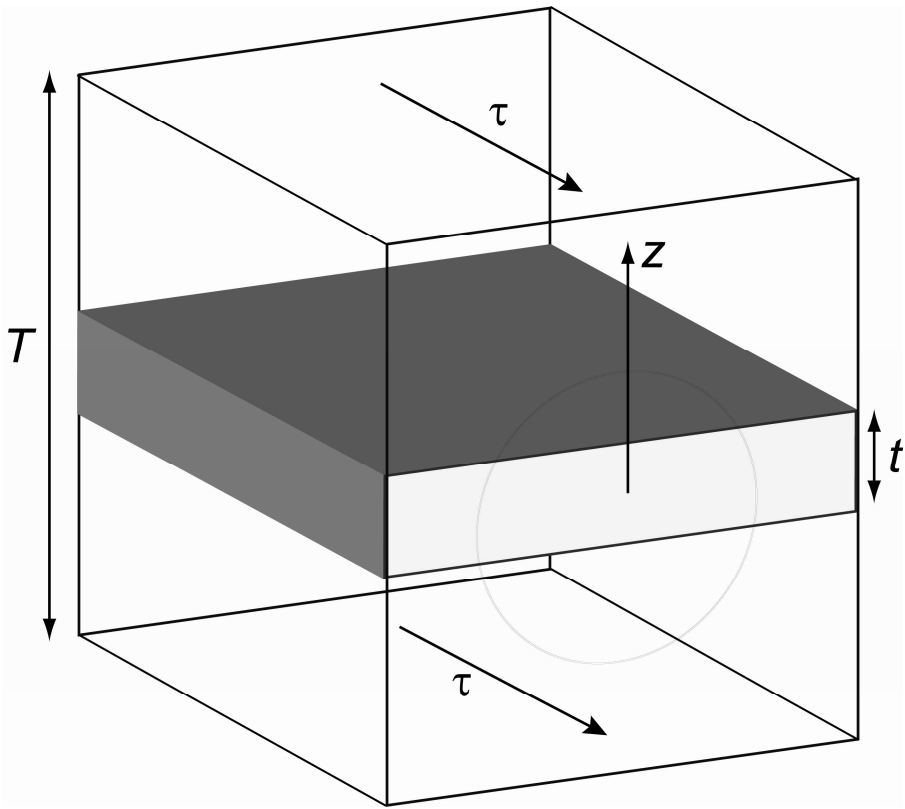

Figure SI.4 Model of a flake within a resin used in shear-lag theory. The shear stress, $\tau$, acts at a distance $z$ from the flake centre.

The behaviour of a discontinuous flake in a matrix can be modelled using shear lag theory in which it is assumed that the flake is surrounded by a layer of resin at a distance, $z$, from the flake centre as show in Fig. SI.4. The resin has an overall thickness of $T$. It is assumed that both the flake and matrix deform elastically and the flake-matrix interface remains intact. If $u$ is the displacement of the matrix in the flake axial direction at a distance, $z$, then the shear strain, $\gamma$, at that position is be given by

$$\gamma = \frac{du}{dz} \quad (SI.3)$$

The shear modulus of the matrix is defined as $G_m = \tau/\gamma$ hence

$$\frac{du}{dz} = \frac{\tau}{G_m} \quad (SI.4)$$

The shear force per unit length carried by the matrix is transmitted to the flake surface though the layers of resin and so the shear strain at any distance $z$ is given by

$$\frac{du}{dz} = \frac{\tau_i}{G_m} \quad (SI.5)$$

This equation can be integrated using the limits of the displacement at the flake surface ($z = t/2$) of $u = u_f$ and the displacement at $z = T/2$ of $u = u_T$

$$\int_{u_f}^{u_T} du = \left(\frac{\tau_i}{G_m}\right)\int_{t/2}^{T/2} dz \quad (SI.6)$$

hence $\quad u_T - u_f = \left(\frac{\tau_i}{2G_m}\right)(T-t) \quad (SI.7)$

It is possible to convert these displacements into strain since the flake strain, $e_f$ and matrix strain, $e_m$, can be approximated as $e_f \approx du_f/dx$ and $e_m \approx du_T/dx$. It should be noted that this shear-lag analysis is not rigorous but it serves as a simple illustration of the process of stress transfer from the matrix to a flake in a graphene-flake composite. In addition, $\tau_i$ is given by Equation (SI.2) and so differentiating Equation (SI.7) with respect to $x$ leads to

$$e_f - e_m = \frac{tT}{2G_m}\left(\frac{d^2\sigma_f}{dx^2}\right) \quad (SI.8)$$

since $T \gg t$. Multiplying through by $E_f$ gives

$$\frac{d^2\sigma_f}{dx^2} = \frac{n^2}{t^2}(\sigma_f - e_m E_f)$$

where $\quad n = \sqrt{\frac{2G_m}{E_f}\left(\frac{t}{T}\right)} \quad (SI.9)$

This differential equation has the general solution

$$\sigma_f = E_f e_m + C\sinh\left(\frac{nx}{t}\right) + D\cosh\left(\frac{nx}{t}\right)$$

where $C$ and $D$ are constants of integration. This equation can be simplified and solved if it is assumed that the boundary conditions are that there is no stress transmitted across the flake ends, i.e. if $x = 0$ in the middle of the flake where $\sigma_f = E_f e_m$ then $\sigma_f = 0$ at $x = \pm l/2$. This leads to $C = 0$ and

$$D = -\frac{E_f e_m}{\cosh(nl/2t)}$$

The final equation for the distribution of flake stress as a function of distance, $x$ along the flake is then

$$\sigma_f = E_f e_m \left[1 - \frac{\cosh(nx/t)}{\cosh(nl/2t)}\right] \quad \text{(SI.10)}$$

Finally it is possible to determine the distribution of interfacial shear stress along the flake using Equation (SI.2) which leads to

$$\tau_i = nE_f e_m \frac{\sinh(nx/t)}{\cosh(nl/2t)}$$

It is convenient at this stage to introduce the concept of flake aspect ratio, $s = l/t$ so that the two equations above can be rewritten as

$$\sigma_f = E_f e_m \left[1 - \frac{\cosh\left(ns\frac{x}{l}\right)}{\cosh(ns/2)}\right] \quad \text{(SI.12)}$$

for the axial flake stress and as

$$\tau_i = nE_f e_m \frac{\sinh\left(ns\frac{x}{l}\right)}{\cosh(ns/2)}$$

for the interfacial shear stress.

It can be seen that the flake is most highly stressed, i.e. the most efficient flake reinforcement is obtained, when the product $ns$ is high. This implies that a high aspect ratio, $s$, is desirable along with a high value of $n$.

**SI.3 Fit of Experimental Data of the Graphene Monolayer to the Shear Lag Analysis**

The experimental data on the variation of graphene strain across the monolayer flake are fitted to the shear lag analysis derived above in Fig. SI.5. It can be seen that the fits of the theoretical shear-lag curves to the strain distribution are sensitive to the value of $ns$ chosen. Likewise the value of interfacial shear stress at the flake ends is very sensitive to the values of $ns$ chosen.

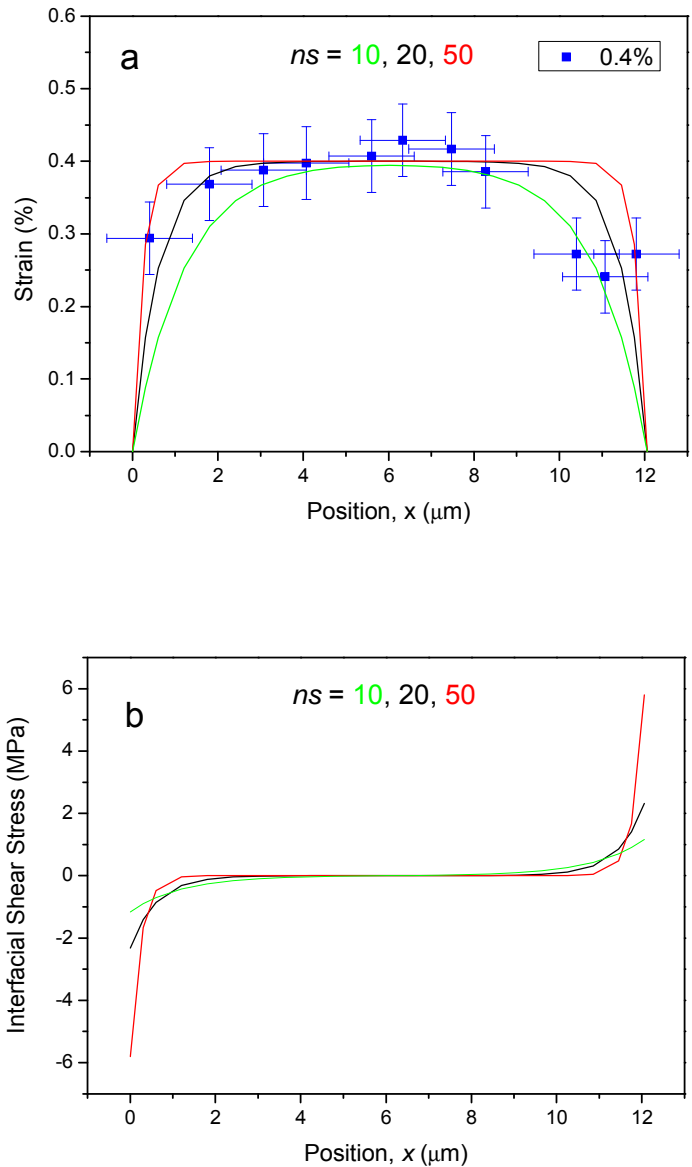

Figure SI.5 **a**. Distribution of strain in the graphene in direction of the tensile axis across a single monolayer at 0.4% strain. The curves are fits of Equ. SI.12 using different values of parameter *ns*. **b**. Variation of interfacial shear stress with position determined from Equ. SI.13 for the values of *ns* used in **a**.

Figure SI.6 shows the fits of Equ. SI.12 to the vertical strain distribution across the graphene monolayer flake as it tapers to a point at *y* = 0. It can be seen that the fits are very sensitive to the value of *ns* employed with the best fit being for *ns* = 20.

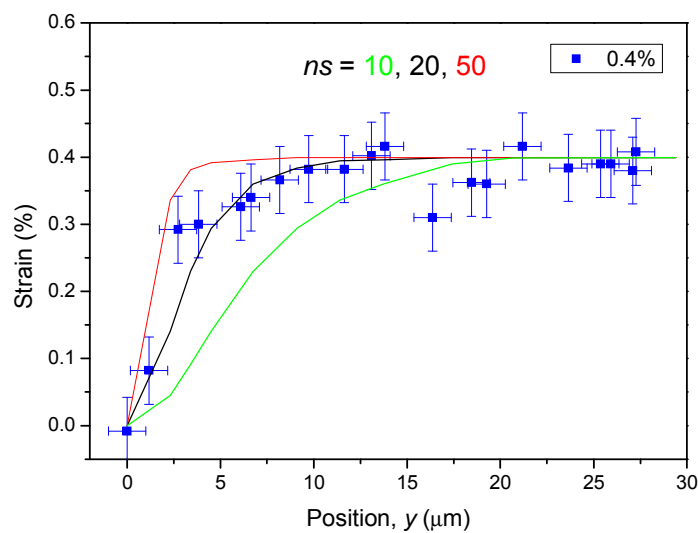

Figure SI.6 Distribution of strain in the graphene in direction of the tensile axis across a single monolayer at 0.4% strain showing the variation of fibre strain with position across the monolayer in the vertical direction. The curves were calculated from Equ SI.12 using different values of *ns*.

**References**

1. Ferrari A. C. *et al.*, Raman spectrum of graphene and graphene layers, *Physical Review Letters*, **97**, 187401 (2006)

2. Kelly, A. & Macmillan, N.H., *Strong Solids*, 3rd Edition, Clarendon Press, Oxford, 1986.

3. Young, R.J. & Lovell, P.A., *Introduction to Polymers*, 3rd Edition, Chapter 24, CRC Press, London, in press